# Accurate Determination of the Charge Transfer Efficiency of Photoanodes for Solar Water Splitting

Dino Klotz, Daniel A. Grave, Avner Rothschild

Department of Materials Science and Engineering, Technion – Israel Institute of Technology, Haifa, Israel.

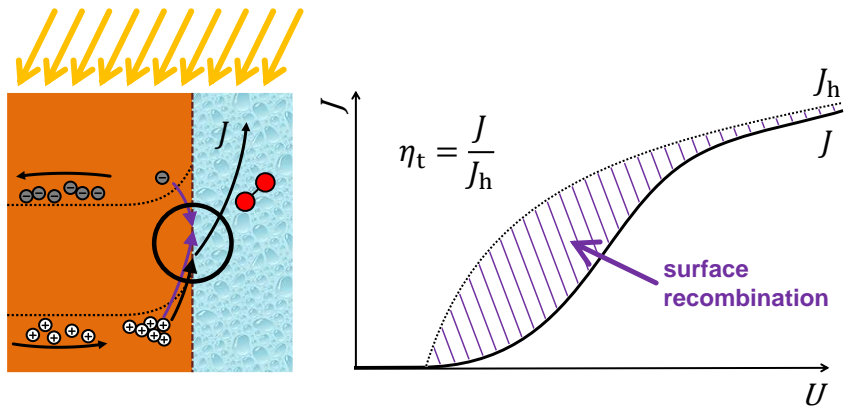

## Abstract

The oxygen evolution reaction (OER) at the surface of semiconductor photoanodes is critical for photoelectrochemical water splitting. This reaction involves photo-generated holes that oxidize water via charge transfer at the photoanode/electrolyte interface. However, a certain fraction of the holes that reach the surface recombine with electrons from the conduction band, giving rise to the surface recombination loss. The charge transfer efficiency, $\eta_t$, defined as the ratio between the flux of holes that contribute to the water oxidation reaction and the total flux of holes that reach the surface, is an important parameter that helps to distinguish between bulk and surface recombination losses. However, accurate determination of $\eta_t$ by conventional voltammetry measurements is complicated because only the total current is measured and it is difficult to discern between different contributions to the current. Chopped light measurement (CLM) and hole scavenger measurement (HSM) techniques are widely employed to determine $\eta_t$, but they often lead to errors resulting from instrumental as well as fundamental limitations. Intensity modulated photocurrent spectroscopy (IMPS) is better suited for accurate determination of $\eta_t$ because it provides direct information on both the total photocurrent and the surface recombination current. However, careful analysis of IMPS measurements at different light intensities is required to account for nonlinear effects. This work compares the $\eta_t$ values obtained by these methods using heteroepitaxial thin-film hematite photoanodes as a case study. We show that a wide spread of $\eta_t$ values is obtained by different analysis methods, and even within the same method different values may be obtained depending on instrumental and experimental conditions such as the light source and light intensity. Statistical analysis of the results obtained for our model hematite photoanode show good correlation between different methods for measurements carried out with the same light source, light intensity and potential. However, there is a considerable spread in the results obtained by different methods. For accurate determination of $\eta_t$, we recommend IMPS measurements *in operando* with a bias light intensity such that the irradiance is as close as possible to the AM1.5 global solar spectrum.

# 1. Introduction

The water photo-oxidation current density, or photocurrent in short, is an important characteristic of semiconductor photoanodes for photoelectrochemical water splitting.[1,2] It measures the rate of the oxygen evolution reaction (OER), and therefore it can be used to evaluate the photoanode's conversion efficiency.[3] Different optical, electrical and electrochemical processes contribute to the photocurrent, starting from absorption of the incident light and ending with oxygen evolution.[4] The line-up of these processes is depicted in Figure 1. They can be assigned with the following efficiencies:

- The *absorption efficiency* ($\eta_{abs}$), also called the light harvesting efficiency, is the fraction of the incident photon flux that is absorbed in the photocatalytic layer and gives rise to photo-generation of minority charge carriers, i.e., holes in photoanodes (productive absorption). The rest of the photons are reflected, transmitted or absorbed in the substrate or other parts of the specimen that do not contribute to the photocurrent (wasted absorption).[5]
- The *charge separation efficiency* ($\eta_{sep}$) is the fraction of the photo-generated holes that reach the surface, whereas the rest of the holes recombine with electrons within the photoanode before reaching the surface (bulk recombination).
- The *charge transfer efficiency* ($\eta_t$), also called the injection efficiency, is the fraction of the holes that give rise to electrochemical reactions out of the holes that have reached the surface. The rest of the holes recombine with conduction band electrons at the surface (surface recombination).
- The *Faradaic efficiency* ($\eta_F$) is the fraction of holes that give rise to the water oxidation reaction and not other side reactions such as corrosion and decomposition. Hematite photoanodes are stable in alkaline aqueous solutions[6] and they often display a Faradaic efficiency of close to 100% for water oxidation.[7]

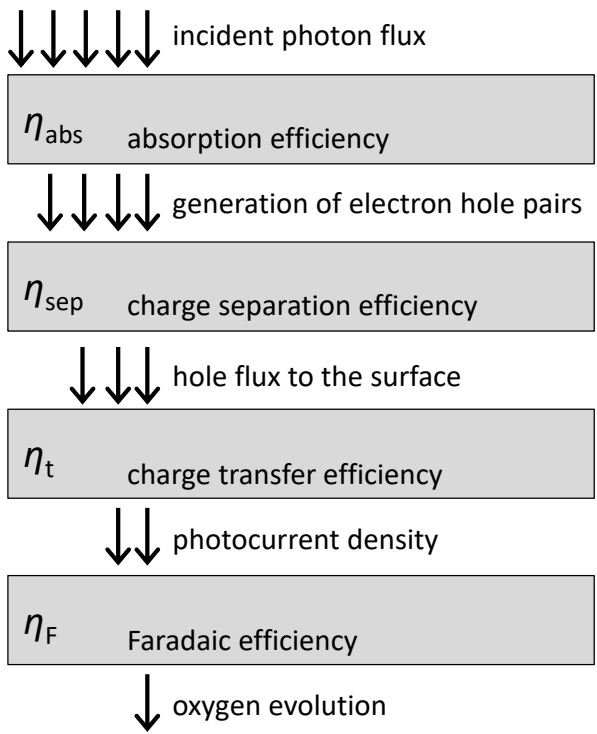

**Figure 1. Line-up of the different processes involved in water photo-oxidation and their efficiencies.**

This line-up of efficiencies can be compiled into equation (1), which links the incident photon flux $\phi_{in}$ with the photocurrent $J$:

$$J = q \cdot \phi_{in} \cdot \eta_{abs} \cdot \eta_{sep} \cdot \eta_t \cdot \eta_F \cdot \qquad (1)$$

Here, $q$ represents the elementary charge. The absorption and Faradaic efficiencies can be evaluated from optical and gas chromatography measurements, respectively, but it is difficult to discern between the charge separation and charge transfer efficiencies. A flawless determination of $\eta_{sep}$ and $\eta_t$ is important to distinguish between surface and bulk recombination losses in order to focus optimization efforts to improve the photoanode efficiency at the most critical segment.[8–10] To address this challenge, different methods have been developed to determine $\eta_{sep}$ and $\eta_t$.[10,11] We will focus on $\eta_t$ in this article, noting that it suffices to determine either $\eta_t$ or $\eta_{sep}$ and then the other parameter can be extracted from the measured photocurrent using equation (1). Different methods to determine $\eta_t$ often yield different results, as demonstrated in Ref. [12] (compare Figures 2c and 6c). Therefore, critical assessment of the applicability of these methods and best-practice guidelines are required for accurate determination of $\eta_t$. Toward this end, this work compares the $\eta_t$ values obtained by different measurement methods, using a hematite photoanode as a case study. The effect of experimental conditions such as the light source and light intensity are also examined.

## 2. Measuring the Charge Transfer Efficiency

The charge transfer efficiency is controlled by processes occurring at the surface of the photoanode. The exact reaction path and intermediate species on the surface are subject to extensive studies and discussions[13–15] that go beyond the scope of this paper. There is, however, a general agreement about two competing paths at the surface of the photoanode, as illustrated in Figure 2. Photo-generated holes that reach the surface can either oxidize water (green arrows in Figure 2) or recombine with electrons from the conduction band (purple arrows).

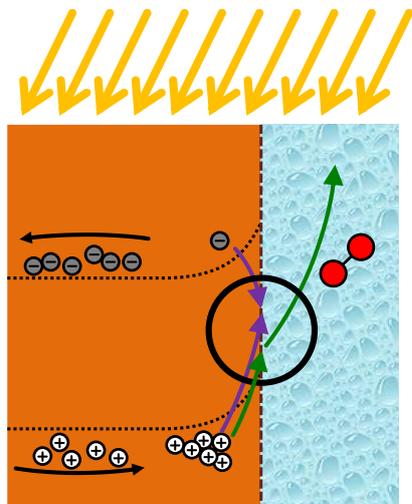

**Figure 2. Schematic illustration of competing paths for the fate of the holes that reach the surface of the photoanode. The desired path (shown by green arrows) leads to water oxidation, whereas the undesired path (purple arrows) leads to surface recombination.**

The following current densities can be assigned to the charge carrier fluxes shown in Figure 2:
- The current density of holes reaching the surface: $J_h = qF_{h,s}$, where $F_{h,s}$ is the flux of holes reaching the surface.
- The surface recombination current density, $J_r$.

From Kirchhoff's current law, the photocurrent $J$ is equal to the sum of $J_h$ and $J_r$:

$$J = J_h + J_r . \qquad (2)$$

$J_h$ is defined as a positive current, whereas $J_r$ is negative and therefore it reduces the photocurrent. The surface recombination current can be expresses in terms of the flux of electrons from the conduction band to the surface (see Figure 2), $J_r = -qF_{e,s}$. Thus, $J_h$ and $J_r$ represent the current densities of holes and electrons, respectively, at the surface. The charge transfer efficiency $\eta_t$ is defined as

$$\eta_t = \frac{J}{J_h} = \frac{J_h + J_r}{J_h} = \frac{J}{J - J_r} \quad . \tag{3}$$

Equation (3) shows that identifying either $J_h$ or $J_r$ in addition to $J$ is sufficient to determine $\eta_t$.

The most commonly used means of characterizing the performance of a photoanode is voltammetry, where the photocurrent $J$ is measured as a function of the applied potential $U$. For benchmarking purposes, voltammetry measurements must be carried out under solar-simulated illumination without sacrificial reagents in the electrolyte. The sweep rate should be slow enough to measure the steady-state photocurrent. A typical voltammogram is illustrated schematically by the black curve in Figure 3a.

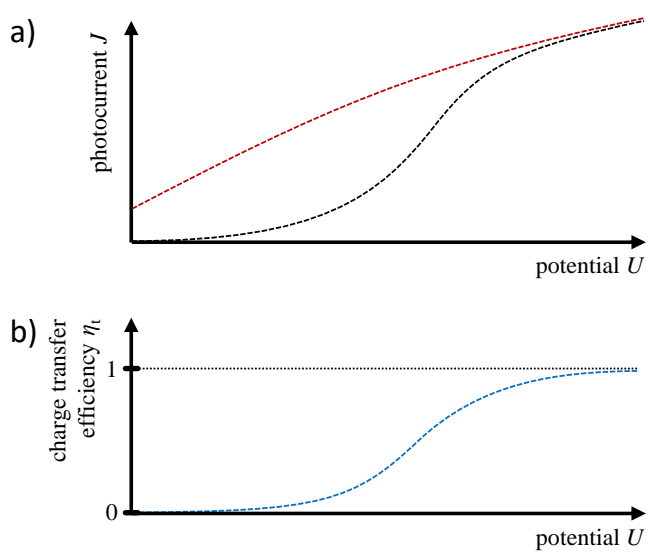

**Figure 3.** Typical shapes of (a) the photocurrent $J$ (black) and hole current $J_h$ (red), and (b) the corresponding charge transfer efficiency, $\eta_t$.

Conventional voltammetry measurements yield the total photocurrent ($J$) and do not carry specific information on the hole current ($J_h$) and surface recombination current ($J_r$).[16] This is because current measurements without selective contacts cannot distinguish between hole and electron currents. Thus, the hole current, shown by the red curve in Figure 3a, has to be determined by other methods, as has been reported elsewhere.[10,12,13,17,18] Figure 3b shows the charge transfer efficiency $\eta_t$, calculated using equation (3) with $J$ and $J_h$ from Figure 3a. $\eta_t$ varies between zero at potentials below the onset potential to one far beyond the onset.[13]

In order to calculate $\eta_t$, $J$ and either $J_h$ or $J_r$ should be determined. Two analytical approaches are used to determine $J_h$ and $J_r$:
1. **Dynamic measurements.** Because of the different time constants with which $J_h$ and $J_r$ are setting in, chopped light measurements (CLM) and intensity modulated photocurrent spectroscopy (IMPS) measurements have the potential to distinguish between the two currents.
2. **Hole scavenger measurements (HSM).** If a hole scavenger is added to the electrolyte, surface recombination can be suppressed and then $J_h$ is obtained from the photocurrent measured with the hole scavenger, whereas the water photo-oxidation current density $J$ is determined without hole scavenger.[10] However, care must be taken because the photocurrent measured with the hole scavenger may not be the same as $J_h$ without the hole scavenger due to spurious effects such as current doubling,[19] or due to changes in the surface potential and space charge characteristics that result from the different surface reaction, as will be discussed in more detail below.

In the following, we compare the $\eta_t$ values of a hematite photoanode determined by CLM, IMPS and HSM. For reliable comparison we use a model hematite photoanode that is known to be stable and to provide reproducible results.[6,20] Toward this end, we use a heteroepitaxial (110) oriented hematite thin-film photoanode (thickness ~30 nm) on Nb-doped $SnO_2$ (NTO) transparent electrode (thickness ~350 nm) deposited on an $a$-plane sapphire substrate. Details of the fabrication process and the photoelectrochemical characteristics of the photoanode can be found elsewhere.[21]

## 2.1 Chopped Light Measurements (CLM)

In the early 1980s, Salvador proposed CLM to distinguish between hole and surface recombination currents and applied the analysis for $TiO_2$ photoanodes.[22] The procedure to obtain $\eta_t$ from CLM is simple. The photoanode is kept in a photoelectrochemical setup with the usual electrolyte in the dark at the test potential. Upon switching the light on, the photocurrent increases almost instantaneously to the maximum value, forming a spike as illustrated in Figure 4. Because of the fast response, this component was called the *"instantaneous hole current"*.[23] Subsequently, the photocurrent decays to a steady-state value. The decay is ascribed to the setting in of the surface recombination current. It occurs with a larger time constant than the hole current. The exact makeup of the processes that lead to the characteristic shape of the transient photocurrent response is still under discussion.[24] It is noteworthy that the spike is not a displacement current that accounts for the transient charging and discharging current through a capacitor. It is not comparable to the current response to a voltage step of a circuit element composed of a resistor and capacitor in parallel. This can be rationalized by the fact that the spike height does not depend on the step time. For a capacitive effect, the spike height would tend to infinity for infinitely small step time. This is not the typical case for CLM of semiconductor photoanodes, where the spike is formed by two separate processes, a fast positive response followed by a slow negative response.[23]

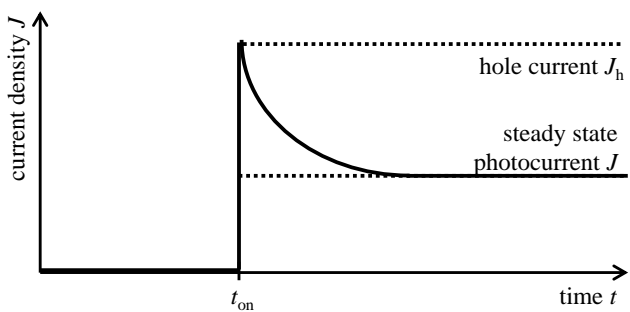

**Figure 4. Qualitative illustration of the transient photocurrent response upon switching the light on at $t_{on}$ (solid line). The dashed lines represent the hole current at the maximum of the peak and the steady-state photocurrent to which the photocurrent relaxes. Reproduced after Refs. [22,23].**

Two values can be extracted from transient photocurrent response curves such as the one illustrated in Figure 4: the hole current, $J_h$, determined by the apex of the spike, and the steady-state photocurrent, $J$. With these values, $\eta_t$ is easily calculated by equation (3). However, the shape of the spike is sensitive to the shutter speed which controls the time it takes to switch the light on and off, so-called the step time. This is demonstrated in Figure 5 that shows two CLM of the same photoanode carried out with different instruments. The black curve was recorded with a mechanical shutter in front of an ABET AAA1.5G solar simulator. The spike is smeared off, showing a transient photocurrent response of several tens of ms upon turning the light on. It is not clear a-priori to what extent the transient characteristics depend on the shutter dynamics. It is noted that step times of 10 ms or more are quite common for mechanical shutters that are often used in conventional solar simulators. Such step times are much slower than the typical photocarrier dynamics of hematite photoanodes,[25,26] suggesting that the transient characteristics in Figure 5 (black curve) were influenced by the slow shutter speed. Indeed, CLM of the same photoanode conducted with a Zahner CIMPS system equipped with a white LED and a fast intensity transients (FIT) module that switches on and off in 1 µs, as confirmed by an internal light detector (not shown here), give rise to a markedly different spike shape, as shown by the red curve in Figure 5. Both black and red curves were recorded at the same photoanode potential (1.35 $V_{RHE}$). The light intensity of the white LED was set to 80 mW/cm² in order to yield a similar flux of photons with energy above the bandgap (2.1 eV) as with a solar simulator. The different spike shapes in Figure 5 reveal that the slow transient response recorded with the solar simulator (black curve) originates from the slow mechanical shutter rather than from the charge carrier dynamics. Consequently, the apex of the spike may be damped by a slow shutter, potentially leading to an error in the $J_h$ estimation. One possibility to overcome this artifact is to fit the recorded transient response to an exponential function in order to estimate the extrapolated apex value. However, the extrapolation depends on the shutter dynamics which must be carefully measured and analyzed for accurate extrapolation. Another consideration is the sampling rate at which the current is recorded. When the carrier dynamics are fast, as in Figure 5 (red curve), the spike is narrow so that it requires sampling rates ≥ 1 kHz to capture the apex of the spike. Apart from instrumental issues such as the shutter speed, some researchers have raised fundamental concerns about CLM since the drastic change between zero to full light intensity may change the surface band bending and space charge characteristics so that the measurement is not conducted in steady-state.[23] For an assessment of the implications of this issue see the results and discussion section.

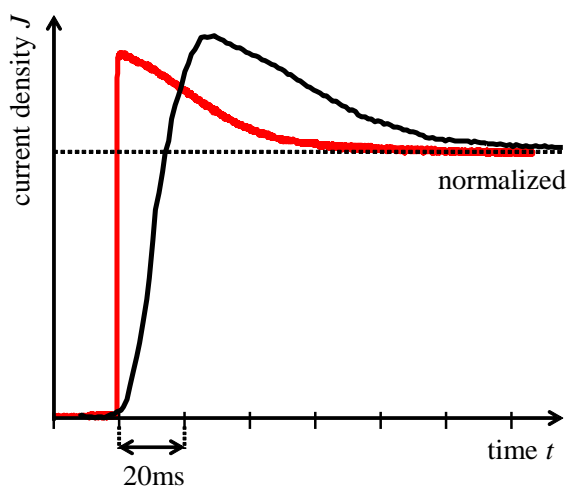

**Figure 5. Comparison of CLM transient photocurrent response curves taken with a mechanical shutter in front of an ABET AAA1.5G solar simulator (black curve), and a Zahner CIMPS system with a white LED controlled by a fast intensity transient (FIT) module (red curve). Both curves were recorded using the same photoanode at a potential of 1.35 $V_{RHE}$. The curves are normalized to yield the same steady-state photocurrent value.**

## 2.2  Intensity Modulated Photocurrent Spectroscopy (IMPS)

IMPS is considered to yield authoritative results for $\eta_t$ because it is based on small-signal light intensity perturbation around the actual operating point of the photoanode. Frequency domain techniques such as IMPS are more robust against distorted excitation signals than time domain techniques such as CLM. Indeed, IMPS is becoming more and more popular as a means to study semiconductor photoanodes.[8,11,12,27] However, the interpretation of the results is not always straightforward. After a brief introduction into the measurement technique itself, we will compare two approaches to calculate $\eta_t$ from IMPS measurements. In IMPS, the photoanode is held at an operating point with fixed bias potential and bias light intensity. A small-signal sinusoidal modulation with a frequency ω is added to the bias light intensity, $I(t)$, and the photocurrent $J(t)$ is measured. A frequency sweep of ω yields the photocurrent admittance spectrum, $Y_{pc}(\omega)$:[26]

$$Y_{pc}(\omega) = \frac{J(\omega)}{I(\omega)}. \tag{4}$$

All IMPS spectra shown here were obtained by PEIS and IMVS measurements and subsequent calculation of the IMPS spectrum as suggested in Ref. [26]. For a detailed introduction to IMPS the reader is referred to the seminal work by Peter and co-workers.[16,24,28] Recently, alternative approaches for IMPS analysis were presented elsewhere.[26]

The most commonly used approach to determine $\eta_t$ from IMPS measurements was introduced by Peter.[28] In brief, it considers the high frequency intersect (HFI) and low frequency intersect (LFI or $Y_{pc}(0)$) of the IMPS spectrum with the real axis, as depicted in Figure 6. The diameters of the lower and upper semicircles in the IMPS spectrum can be related, assuming a simple linear model, to the rate constants for charge transfer and recombination, $k_t$ and $k_r$, respectively. With the quantities $k_t = d_t$ and $k_r = d_r$ obtained from the IMPS spectrum (as shown in Figure 6), $\eta_t$ is calculated using the following equation:[17]

$$\eta_t = \frac{k_t}{k_t + k_r} = \frac{d_t}{d_t + d_r} \tag{5}$$

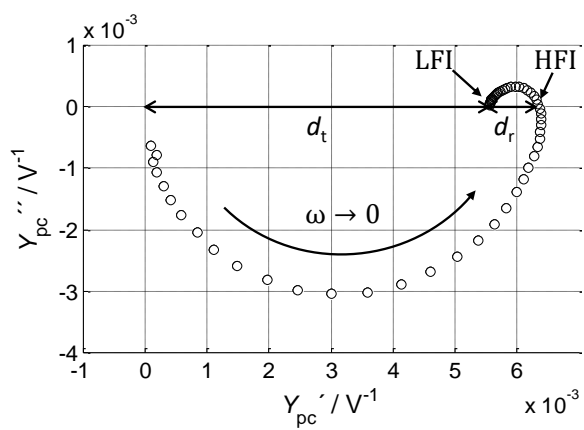

**Figure 6.** A typical IMPS spectrum of the model hematite photoanode, measured at 1.35 $V_{RHE}$ with a white LED providing a bias light intensity of 100 mW/cm². The low frequency intersect (LFI or $Y_{pc}(0)$) and high frequency intersect (HFI) with the real axis are marked. The lower and upper semicircles are called $Y_{pc}^+(\omega)$ and $Y_{pc}^-(\omega)$, respectively.[26]

A more accurate way to determine the magnitude of the positive and negative semicircles, $Y_{pc}^+(0)$ and $Y_{pc}^-(0)$, respectively, was proposed in Ref. [26] by fitting the IMPS spectrum to an equivalent circuit model (ECM), instead of extracting the $d_t$ and $d_r$ diameters directly from the IMPS spectrum, as in Figure 6. At the position around the HFI, the opposite imaginary parts of the upper and lower semicircles diminish each other, causing the HFI to be shifted to the left. The shift becomes considerable if the time constants of the semicircles are similar. This point was already raised in Ref. [23], where the authors stated that the time constants must be separated by at least two orders of magnitude, otherwise there is a systematic error in the value of the HFI. However, we have previously demonstrated that small errors in the determination of the HFI may arise even when the time constants are separated by more than two orders of magnitude.[26] These errors can be rectified by fitting the IMPS spectrum to an ECM. The simple ECM illustrated in Figure 7 suffices for accurate determination of the $Y_{pc}^+(0)$ and $Y_{pc}^-(0)$ values. From this fitting we extract

$$k_t = Y_{pc}^+(0) + Y_{pc}^-(0) = Y_1 + Y_2 + Y_3, \qquad (6)$$

$$k_r = -Y_{pc}^-(0) = -Y_3, \qquad (Y_3 < 0). \qquad (7)$$

However, it was shown in Ref. [26] that this approach does not account for nonlinear response of the photocurrent to the bias light intensity. Such nonlinear response may arise from nonlinear relationship between the recombination current and the bias light intensity, or other effects. To account for those nonlinear effects, a rigorous IMPS analysis method was proposed,[26] which will be briefly introduced in the following.

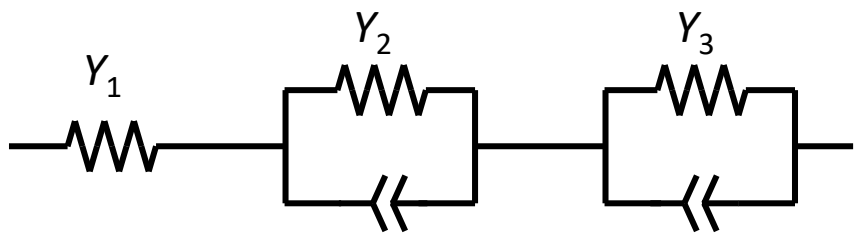

**Figure 7.** A simple ECM for fitting IMPS spectra to obtain accurate $Y_{pc}^+(0)$ and $Y_{pc}^-(0)$ values. Exemplary fits and the corresponding residuals are shown in Figures S2 to S4.

A rigorous IMPS analysis should account for possible nonlinear response of the currents ($J$, $J_h$ and $J_r$) to the bias light intensity ($I$). However, IMPS probes the linear behavior at the operating point by measuring the response to small-signal perturbations around this point. In order to account for nonlinear $J$-$I$ behavior, it was proposed to measure IMPS spectra at different bias light intensities.[26] Thus, the $Y_{pc}^+(0)$ and $Y_{pc}^-(0)$ values are obtained from every IMPS spectrum (at different light intensities) by ECM fitting, as explained before. Then, the currents $J$, $J_h$ and $J_r$ are reconstructed with a simple polynomial fitting that takes into account the $Y_{pc}^+(0)$, $Y_{pc}^-(0)$, $Y_{pc}(0) = Y_{pc}^+(0) + Y_{pc}^-(0)$ values and the absolute value of the photocurrent at the operating point in which the IMPS spectrum was measured. More details on this approach can be found elsewhere, see section S5 in the supporting information of Ref. [26].

Figure 8 shows a diagram where the rigorous IMPS analysis was applied to two IMPS measurements with a white LED providing bias light intensities of 50 and 100 mW/cm². It shows $J$, $J_h$ and $J_r$ as a function of the bias light intensity for a bias potential of 1.35 $V_{RHE}$. While $J_h$ is quite linear, nonlinearities in both $J$ and $J_r$ are observable, although not very pronounced. The dashed lines serve as guides for the eye featuring linear $J$-$I$ curves. The deviation from these lines display small nonlinearities in $J$ and $J_r$. Because of the rather small nonlinearity, measurements at two bias light intensities suffice for accurate determination of all the currents.

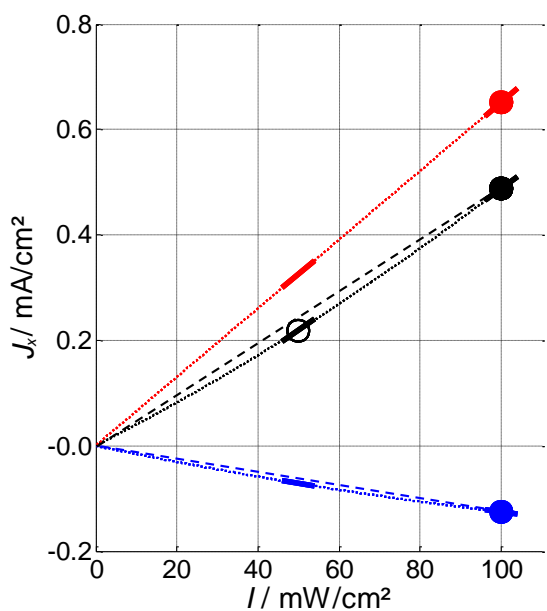

**Figure 8.** Rigorous IMPS analysis of the model hematite photoanode (measured at 1.35 $V_{RHE}$ with a white LED), plotting the current densities $J$, $J_h$ and $J_r$ (shown by black, red and blue, respectively) as a function to the bias light intensity $I$. The thick lines at $I$ = 50 and 100 mW/cm² show the local derivatives obtained from the IMPS spectra ($Y_{pc}(0)$, $Y_{pc}^+(0)$ and $Y_{pc}^-(0)$). Dotted lines: fitted $J$–$I$ curves; black circles: operating points for IMPS measurements; solid circles: fitted values for $J$, $J_h$ and $J_r$ at 100 mW/cm². The dashed lines are guides for the eye featuring linear $J$-$I$ curves. Corresponding plots for the other LEDs are provided in Figures S7 to S10.

## 2.3   Hole Scavenger Measurements (HSM)

HSM are aimed at removing the barrier for charge transfer of photo-generated holes to slow intermediate states of the OER. By doing so, surface recombination is suppressed and the hole current to the surface is uncovered.[10] This is achieved by adding to the electrolyte sacrificial reductant reagents that react fast with the holes arriving at the surface, faster than the rate of surface recombination. Sacrificial reagents serving as hole scavengers include $H_2O_2$,[10] KI (iodide),[15] and methanol.[29,30] In HSM, voltammetry measurements are taken with and without the hole scavenger in the same settings as the conventional measurements. The $J$-$U$ voltammogram measured with hole scavenger lies above the one measured without it, as demonstrated in Figure 9. If the prerequisites for HSM analysis prevail, as discussed in Ref. [10], the difference between the two curves yields the surface recombination current.

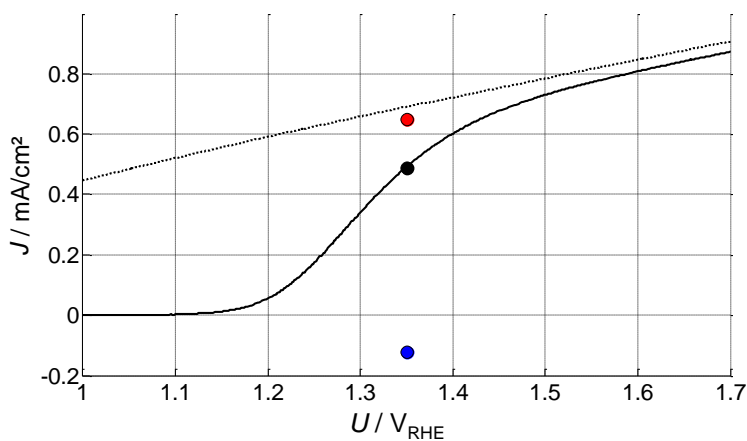

**Figure 9.** Voltammetry measurements of the model hematite photoanode taken with (dotted line) and without $H_2O_2$ (solid line) under illumination of a white LED at 100 mW/cm². The black dot indicates the measured photocurrent obtained from IMPS measurement and the red and blue dots indicate $J_h$ and $J_r$ as determined by the rigorous IMPS analysis at a bias potential of 1.35 $V_{RHE}$ (see Figure 8).

HSM are simple and easy to carry out, but care must be taken to avoid false analysis and potential errors. Due to the high reactivity of some sacrificial reagents such as $H_2O_2$ and the volatility of other reagents such as methanol, HSM must be performed with fresh solutions. Careful consideration is required in cases where the HSM are suspected not to reflect the pertinent currents without the presence of the hole scavenger. In particular, photoanodes that display unusually high currents in the presence of the hole scavenger must be analyzed very carefully. For example, high currents that have nothing to do with water oxidation are often observed when metals such Pt are exposed to electrolytes containing $H_2O_2$ as a result of spontaneous $H_2O_2$ decomposition.[31] This phenomenon may occur even without Pt co-catalysts at the surface. For instance, polycrystalline hematite photoanodes on platinized silicon wafers have shown unstable dark current behavior, probably as a result of pinholes or micro-cracks acting as shunts between the Pt layer and the electrolyte.[32] Therefore, high quality films devoid of pinholes and micro-cracks are necessary for reliable HSM when Pt is used as a back contact.[32] Another effect that potentially obstructs reliable HSM analysis is current doubling, which was observed in $CuWO_4$ photoanodes using $H_2O_2$ as a hole scavenger.[19] It is noted that current doubling goes along with an additional semicircle in the IMPS spectrum, which normally is not observed for hematite photoanodes in operation with $H_2O_2$.[26] As discussed in Ref. [10], hole scavengers alter the electrochemical reaction occurring at the photoanode/electrolyte interface. This may modify the surface charge, which may in turn modify the hole current due to changes in the surface band bending and space charge width. Therefore, HSM may potentially lead to inaccurate estimation of $\eta_t$. As noted in Ref. [10], a necessary indication for reliable determination of $\eta_t$ by HSM is the convergence of the $J$-$U$ voltammograms obtained with and without the hole scavenger at high potentials, where $\eta_t$ is expected to reach 100%. If the voltammograms do not converge at high potentials, as observed for instance in the Zn-doped and undoped hematite photoanodes reported in Ref. [33], the analysis of HSM according to the procedure proposed in Ref. [10] may lead to a false estimation of $\eta_t$. Careful examination of the HSM shown in Figure 9 reveals two things. First, there is a small difference in the hole current obtained by HSM and IMPS analyses at 1.35 $V_{RHE}$. Second, the $H_2O_2$ photocurrent does not fully converge with the photocurrent without $H_2O_2$ for potentials above 1.6 $V_{RHE}$, giving rise to a gap of about 0.05 mA/cm² between the respective curves. The gap indicates that the HSM analysis slightly underestimates $\eta_t$ in this particular case.

## 3. Results and Discussion

In order to demonstrate the main obstacles for accurate determination of $\eta_t$ and the dependence on measurement conditions such as the light source and light intensity, we compare the results obtained for our model hematite photoanode using different methods and different measurement conditions. We begin with CLM. Figure 10 presents transient photocurrent responses obtained by CLM at a potential of 1.35 $V_{RHE}$ using different light sources and different light intensities as detailed in Table 1. All the transient responses were recorded using a Zahner CIMPS system equipped with a FIT module with a step time of ~1 μs, except for the yellow one that was recorded with a solar simulator equipped with a mechanical shutter with a step time of several tens of ms. The different characteristics of the transient responses demonstrate the qualitative differences between the respective CLM. The apex (peak) values, steady-state values and relaxation times are different for different light sources, light intensities, and step times. The implication on the $\eta_t$ values extracted from these measurments is summarized in Table 1 (6$^{th}$ column). The spread in the $\eta_t$ values ranges from 20% for the orange LED operated at 4.5 mW/cm$^2$ to 77.5 % for the blue LED operated at 80 mW/cm$^2$, demonstrating the sensitivity to the light source and light intensity. The transient responses recorded with the orange LED display considerably lower signal-to-noise ratio than the rest of the responses (see the inset in Figure 10), indicating that they should be considered with extra caution. It is noteworthy that the $\eta_t$ values extracted from these responses are much lower than for the other light sources (see Table 1).

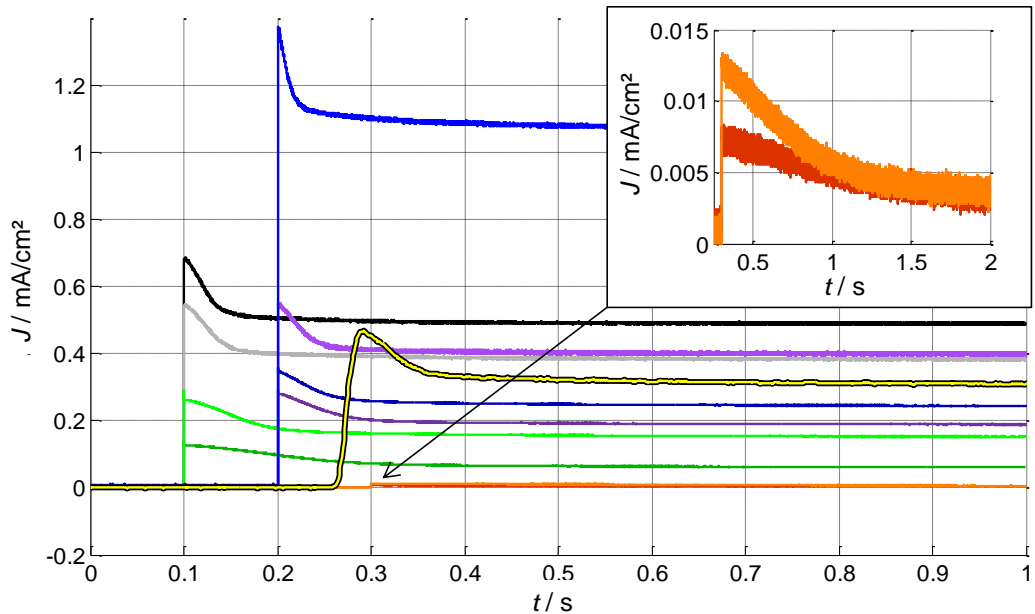

**Figure 10. Transient photocurrent responses obtained by CLM of the model hematite photoanode, recorded at a potential of 1.35 V$_{RHE}$ using different light sources and light intensities as indicated by the color code presented in Table 1. The inset shows a magnification of the transient responses obtained with the orange LED at light intensities of 4.5 and 9 mW/cm$^2$ (dark and light orange curves, respectively).**

**Table 1. Compilation of the $\eta_t$ values obtained using different methods, light sources and light intensities. The first column indicates the color code in Figure 10 and 11.**

| Color | Light Source | Photon Energy / eV | Bias Light Intensity / mW/cm² | Photon Flux / (cm²s)$^{-1}$ | CLM | IMPS Model Approach | Rigorous IMPS Analysis | HSM |
|---|---|---|---|---|---|---|---|---|
| | Solar simulator | | | | 64.9% | | | |
| ○ | White LED (4300 K) | | 50 | | | 80.7% | 67.9% | |
| | White LED (4300 K) | | 80 | | 70.5% | | 72.3% | 67.7% |
| | White LED (4300 K) | | 100 | | 71.5% | 85.9% | 75.3% | 70.6% |
| | Ultraviolet LED (395 nm) | 3.14 | 10 | 1.99 · 10$^{16}$ | 66.3% | 72.5% | 63.3% | |
| | Ultraviolet LED (395 nm) | 3.14 | 20 | 3.98 · 10$^{16}$ | 72.6% | 78.3% | 68.7% | |
| | Blue LED (449 nm) | 2.76 | 20 | 4.52 · 10$^{16}$ | 69.7% | 72.9% | 67.2% | |
| | Blue LED (449 nm) | 2.76 | 80 | 1.81 · 10$^{17}$ | 77.5% | 81.1% | 75.7% | |
| | Green LED (530 nm) | 2.34 | 15 | 4.00 · 10$^{16}$ | 50.0% | 67.5% | 56.0% | |
| | Green LED (530 nm) | 2.34 | 30 | 8.00 · 10$^{16}$ | 58.0% | 75.9% | 63.6% | |
| | Orange LED (590 nm) | 2.10 | 4.5 | 1.34 · 10$^{16}$ | 33.3% | 25.8% | 22.2% | |
| | Orange LED (590 nm) | 2.10 | 9 | 2.68 · 10$^{16}$ | 20.0% | 18.7% | 20.3% | |

Shifting gears to IMPS measurements, they also yield different spectra for different light sources and bias light intensities, as shown in Figure 11. The $\eta_t$ values calculated by fitting the IMPS spectra to the ECM in Figure 7 to obtain the $Y_{pc}^+(0)$ and $Y_{pc}^-(0)$ values, from which $\eta_t$ was calculated using equations (6) and (7), are listed in Table 1 (7$^{th}$ column). These results display similar trends for the dependence of $\eta_t$ on the light source and bias light intensity as in the CLM results. However, despite the qualitative agreement in the trends, the actual $\eta_t$ values are quite different, as can be seen by comparing the values listed in the respective columns in Table 1. The IMPS spectra obtained using the orange LED are very noisy (see magnified spectra in Figure S1 in the Supporting Information), similarly to the noisy CLM for this light source (see inset in Figure 10). However, the fitting results for the orange LED that are also shown in Figures S4 and S5 suggest that the relevant information can be deduced accurately from IMPS spectra despite the noise, which is a known advantage of frequency domain techniques. The IMPS spectra in Figure 11 were also analyzed by the rigorous IMPS analysis method using two measurements at different light intensities, as shown in Figure 8. The results are given in Table 1 (8$^{th}$ column). The $\eta_t$ values obtained by the rigorous IMPS analysis are considerably lower than the respective values obtained by the IMPS model approach.

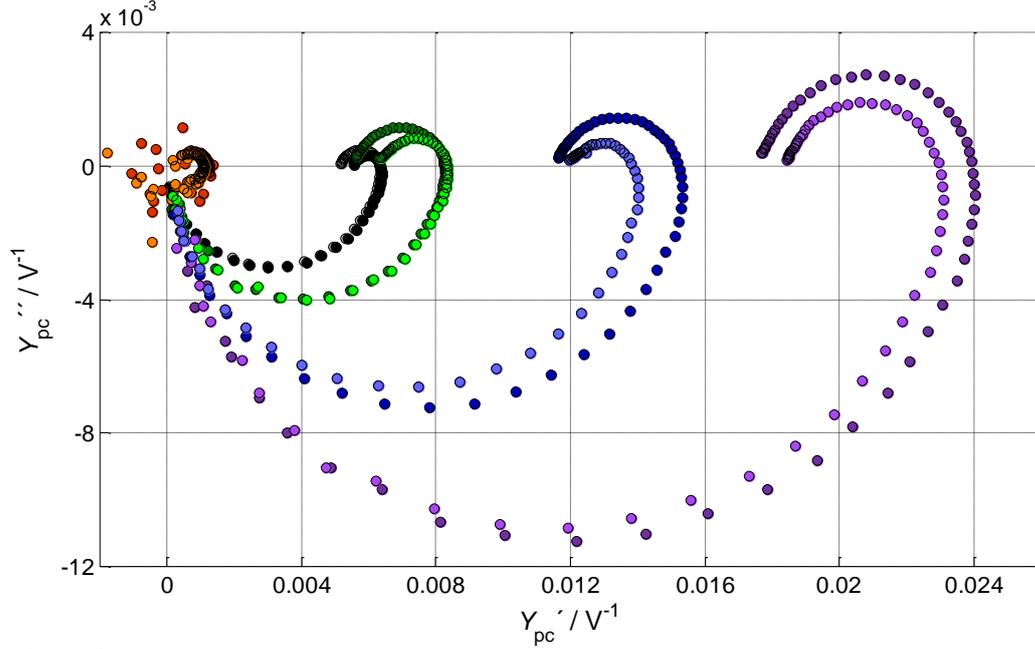

**Figure 11. IMPS spectra obtained using different LEDs operated at different bias light intensities. The resepctive color code and $\eta_t$ values are given in Table 1. Zoom-in of the spectra obtained using the white and orange LEDs is provided in Figure S1.**

Lastly, Table 1 also shows two results for HSM, measured with the white LED (9[th] column). The measurement with the light intensity of 100 mW/cm² was already shown in Figure 9. The $\eta_t$ values obtained by HSM are lower than the rigorous IMPS analysis, yielding the lowest values in this comparative study.

Table 2 presents a statistical analysis of the correlation between $\eta_t$ values obtained by different methods (from Table 1), comparing the CLM, IMPS model approach and HSM results to the rigorous IMPS analysis. The analysis shows a small (-1.5%) systematic deviation between the CLM and the rigorous IMPS analysis, whereas the IMPS model approach yields a large systematic deviation of +11.6% with respect to the rigorous analysis approach. The standard deviation (σ) with respect to the rigorous IMPS analysis is larger for the CLM (0.057) than for the IMPS model approach (0.042). For the HSM there are not enough data points for a meaningful statistical analysis.

**Table 2. Correlation analysis of the results presented in Table 1 with respect to the rigorous IMPS analysis. The fitting diagrams are shown in Figure S6.**

|  | Slope (Expected value) | Systematic error (Deviation from the expected value) | σ (Standard deviation) | $R^2$ (Coefficient of determination) |
|---|---|---|---|---|
| CLM | 0.985 | -1.5% | 0.057 | 0.910 |
| IMPS model approach | 1.116 | +11.6% | 0.042 | 0.875 |
| HSM | 0.960 | -4.0% | 0.019 | 0.995 |

Based on these results, we recommend the rigorous IMPS analysis as the most accurate and most robust method to determine $\eta_t$. This recommendation is based on the attributes of a robust frequency domain technique with a small-signal perturbation that considers IMPS spectra taken at at several bias light intensities to account for possible nonlinear *J-I* behavior. This is the main reason why the $\eta_t$ values obtained by the IMPS model approach are larger (by 11.6%, on average) than those of the rigorous IMPS analysis approach. The results presented in Figure 8 display a sub-linear dependence of the surface recombination current with the bias light intensity, which may account for the overproportionate increase of the photocurrent with concentrated light as reported elsewhere.[34] This is supported by the fact that $Y_{pc}^-(0)$ changes with the bias light intensity to a larger extent than $Y_{pc}^+(0)$, as can be seen in Figure 11. Consequently, nonlinear *J-I* responses are observed for *J* and $J_r$, whereas $J_h$ increases linearly with the light intensity, as shown in Figure 8. The IMPS analysis is preferred over the HSM method because it probes the water photo-oxidation reaction rather than another reaction (i.e., the photo-oxidation of the hole scavenger) that may modify surface properties such as surface charge, surface potential and surface band bending that may influence the hole current.

One of the most important observations in this study is the influence of the light source on the $\eta_t$ values (see Table 1). Therefore, we advise carrying out the analysis with a white light source whose spectrum matches the solar spectrum as much as possible in the spectral range wherein the photoanode is active. A possible way around spectral mismatch between the light source and the solar spectrum would be to use a solar simulator as a bias light source and superimpose a small-signal perturbation by a white LED.

Another interesting observation is the increased hole current in CLM and the larger lower semicircle ($Y_{pc}^+(0)$) in IMPS for shorter wavelengths. We have only shown the impact of the light source as a function of the light intensity in this study. The increase would be even more severe if the light intensity was normalized by the photon energy, in order to analyze the dependency of the photocurrent to the respective photon flux. The photon fluxes corresponding to the light sources and light intensities applied in Figure 10 and 11 are also provided in Table 1 (5th column). It seems that holes generated by high energy photons are less prone to bulk recombination. Shorter wavelengths also yield higher $\eta_t$ when comparing equal light intensity for different wavelengths, for example 20 mW/cm² for the ultraviolet and blue LEDs. This is in line with IPCE and APCE measurements that reveal much higher efficiencies for shorter wavelengths.[35] In the future, the rigorous IMPS analysis could provide additional information to IPCE and APCE measurements, namely a separation of bulk and surface effects as a function of wavelength and light intensity.

## 4. Conclusions

The charge transfer efficiency $\eta_t$ of semiconductor photoanodes for water photo-oxidation is an elusive property that enables to distinguish between bulk and surface recombination processes. However, accurate determination of $\eta_t$ is not straightforward and there is a large spread between different measurement techniques, analysis methods, instruments, light sources and light intensities. Thus, in order to compare the charge transfer efficiency of different samples they must be measured by the same method, under the same experimental conditions, and the results should be analyzed in the same way. Hole scavenger measurements (HSM) are often used to estimate the charge transfer efficiency, but in some cases they yield inflated photocurrents that may give rise to errors in the analysis. Chopped light measurements (CLM) often display slow shutter response, especially with mechanical shutters that are often used with solar simulators that may give rise to errors. The commonly used simple analysis of intensity modulated photocurrent spectroscopy (IMPS) overestimates $\eta_t$ at high bias light intensities because it does not take into account nonlinear response of the photocurrent with respect to light intensity. Rigorous analysis of IMPS spectra measured at several light intensities can overcome this fault and therefore it is regarded as the most reliable method for accurate determination of $\eta_t$. In view of the strong dependence of the charge transfer efficiency on the light source and bias light intensity, we strongly recommend using light sources with spectral output and intensity as close as possible to the solar spectral irradiance standard (AM1.5 Global) in the wavelength range where the photoanode is active. A good indication for this is a steady state photocurrent comparable to the one measured with a calibrated solar simulator.

## Acknowledgements

The research leading to these results has received funding from the European Research Council under the European Union's Seventh Framework Programme (FP/2007–2013)/ERC Grant Agreement no. 617516. D. A. Grave acknowledges support by Marie-Sklodowska-Curie Individual Fellowship no. 659491. The results were obtained using central facilities at the Technion's Hydrogen Technologies Research Laboratory (HTRL), supported by the Adelis Foundation and by the Solar Fuels I-CORE program of the Planning and Budgeting Committee and the Israel Science Foundation (Grant no. 152/11).